# Robust estimation of the parameters of a disturbed non-stationary Gaussian process

**Sergio Frasca[1,2] and Pia Astone[1]**

1 - INFN – Roma 1

2- Dipartimento di Fisica Università Sapienza - Roma

**Abstract.** A typical problem in the detection of the gravitational waves in the data of gravitational antennas is the non-stationarity of the Gaussian noise (and so the varying sensitivity) and the presence of big impulsive disturbances. In such conditions the estimation of the standard deviation of the Gaussian process done with a classical estimator applied after a "rough" cleaning of the big pulses often gives poor results.

We propose a method based on a matched filter applied to an AR histogram of the absolute value of the data.

## 1. Introduction

The data of a gravitational antenna can be modeled as a Gaussian process, sometimes with slowly varying parameters, with some pulses added (normally due to local disturbances). These disturbances are normally big unmodeled pulses that occur randomly and affect the tails of the data samples distribution; often it turns out that the disturbing pulses have a Laplace distribution, but normally we don't know this distribution. In figure 1 there is a plot of a typical disturbed distribution.

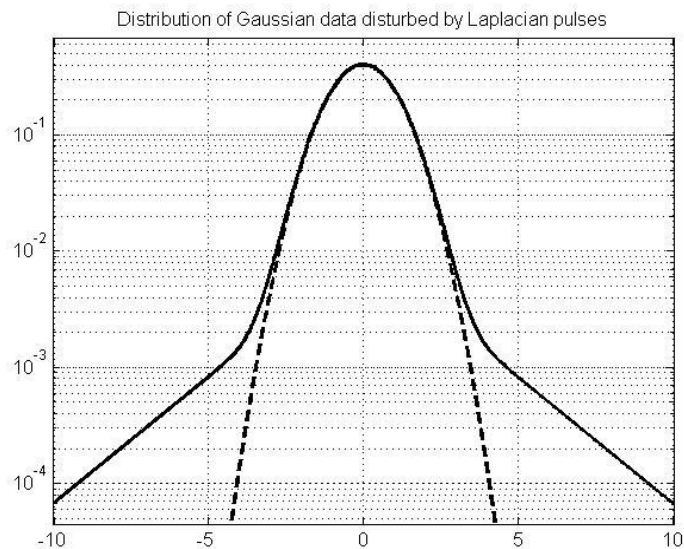

**Figure 1**

The same model applies to the output of any linear filter applied to that data.

# Robust estimation for Gaussian process parameters

A typical problem is the estimation of the (varying) standard deviation of the Gaussian process. This is of fundamental importance in order to optimally detect (and subtract) the disturbing pulses.

The classical estimator for the standard deviation

(1) $$\tilde{\sigma} = \sqrt{\frac{1}{N-1}\sum_{i=1}^{N} x_i^2}$$

doesn't work to estimate the Gaussian process $\sigma$ because of the presence of the disturbances, so we need a more robust one, that gives reasonably correct results as independently as possible from the pulse shape and distribution.

In this paper we present two such estimators, analyze them in the stationary case and then propose an adaptive scheme for the use in the non-stationary case.

## 2. The transformation of the data

Consider the data samples $x_i$ of the type described in the preceding section. Suppose for simplicity that they are independent each other and 0-mean. Let us compute the samples

(2) $$z_i = \log(|x_i|)$$

If the x data were simply gaussian, with 0 mean and standard deviation $\sigma$, the z data should have distribution

(3) $$f(z) = \sqrt{\frac{2}{\pi}} \cdot e^{z-\log\sigma} \cdot \exp\left(\frac{-e^{2\cdot(z-\log\sigma)}}{2}\right)$$

In figure 2 there is the graph.

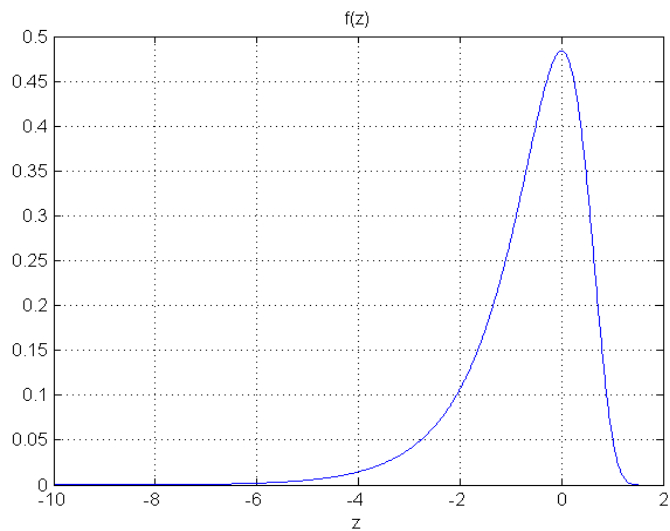

**Figure 2**

## 3. A first estimator

The logarithm reduces the weight of the disturbances, so also simply using the mean value of z we can estimate the value of the σ of the Gaussian process, in a more robust way. In fact, in the case of unit variance, because

$$E[z] = -\frac{2}{\pi} \tag{4}$$

we can estimate the standard deviation of the Gaussian process as

$$\tilde{\sigma} = \exp\left(\bar{z} + \frac{2}{\pi}\right) \tag{5}$$

where $\bar{z}$ is the average value of the $z_i$ samples.

## 4. A second, better estimator

Note that, because of the logarithm, the shape of the distribution is exactly the same independently by the amplitude of the Gaussian process and only the position changes.
The idea is: search with a **matched filter** (e.g. Papoulis 1977) with that shape on the **histogram**. The estimated amplitude is the position of the matching (the "instant"). We recall that a matched filter is a linear running filter that does the convolution of the input data with a template with the shape of the expected pulse.
So, chosen a bin width $\Delta z$, do the histogram of z samples and normalize them by the total number N of the samples, obtaining the values $h_k$ of the histogram. Then, from equation obtain

$$f_k = \sqrt{\frac{2}{\pi}} \cdot e^{k \cdot \Delta z} \cdot \exp\left(\frac{-e^{2 \cdot k \cdot \Delta z}}{2}\right) \tag{6}$$

with k variable in a suitable range such that covers the z range, and do the operation

$$y_k = \sum_l h_{k+l} \cdot f_l \tag{7}$$

From the value k for which $y_k$ is maximum, we estimate the value of the standard deviation as

$$\tilde{\sigma} = \exp(k \cdot \Delta z) \tag{8}$$

From the amplitude of the output of the spectrum we can derive the percentage of non-disturbed Gaussian data and, conversely, the fraction of the "disturbed" time. In fact

$$y_{max} \approx \int_{-\infty}^{\infty} f^2(z) \cdot dz = \frac{1}{\pi} \tag{9}$$

so, if we have a non-disturbed Gaussian process the output of the matched filter is $1/\pi$; if the Gaussian process is disturbed, the fraction of disturbed samples is

Robust estimation for Gaussian process parameters

(10) $$p_{dist} \approx 1 - \pi \cdot y_{max}$$

A problem is that the precision that we can attain on the estimated $\sigma$ is limited by the value of $\Delta z$. A much better precision can be achieved by the following procedure:

- convolve the normalized histogram and the theoretical distribution
- take the fft of the result
- enlarge with zeroes the fft (say, 100 times more)
- take the inverse fft

The result is an enhancement of resolution (and precision) of the same factor (say, 100 times more).

## 5. Results

Here are the results of some simulations. For each simulation we compute the standard deviation estimation by the classical method (eq. (1); solid line), the mean log method (eq. (5); dashed line) and the matched filter method (eq. (8), enhanced by the fft procedure; dot line). For each simulation we present the absolute value of the mean estimation error and the estimation standard deviation (computed, for each point, with 100 simulations), vs the number of the analyzed samples. In all cases the bin width of the histogram is 0.05 and the fft enlargement factor is 100, so the theoretical matched filter resolution is about 1/2000.

In figures 3 and 4 is the case of a Gaussian process of zero mean and unitary variance with 1 % of disturbances, with Gaussian distribution with 0 mean and variance 100.

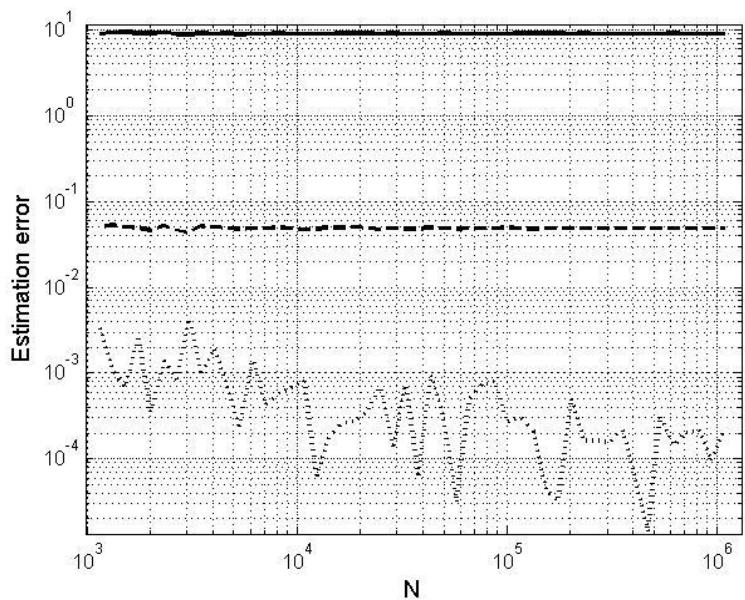

**Figure 3**

Robust estimation for Gaussian process parameters

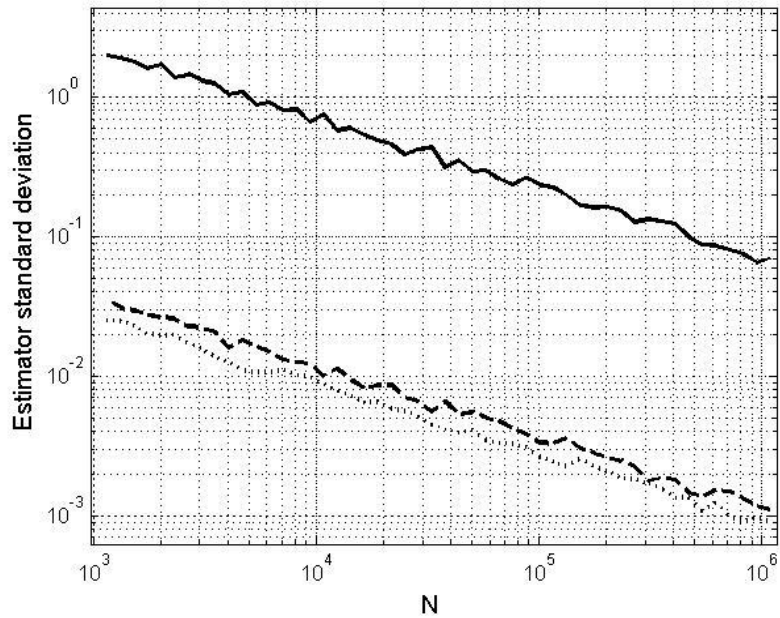

**Figure 4**

We can see that the best estimator is the "matched filter", both for the error and the standard deviation. For high values of N, we reach the theoretical resolution error.

In figures 5 and 6 there is the case of a Gaussian process of zero mean and unitary variance with 40 % of disturbances, with Gaussian distribution with 0 mean and variance 100:

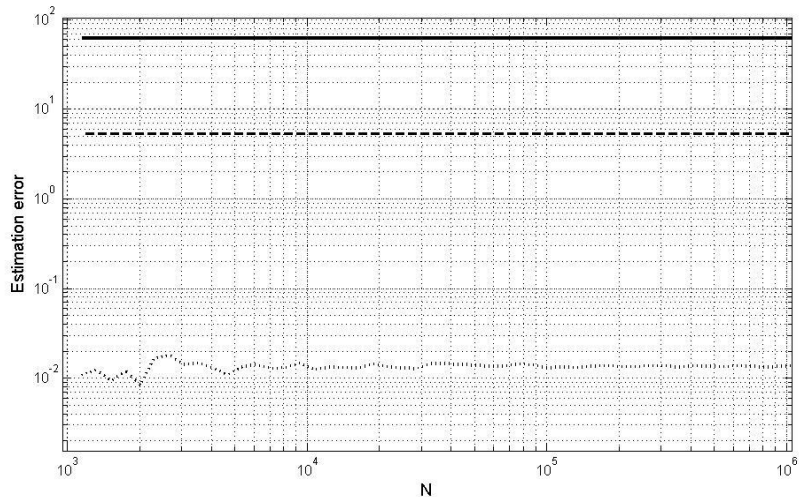

**Figure 5**

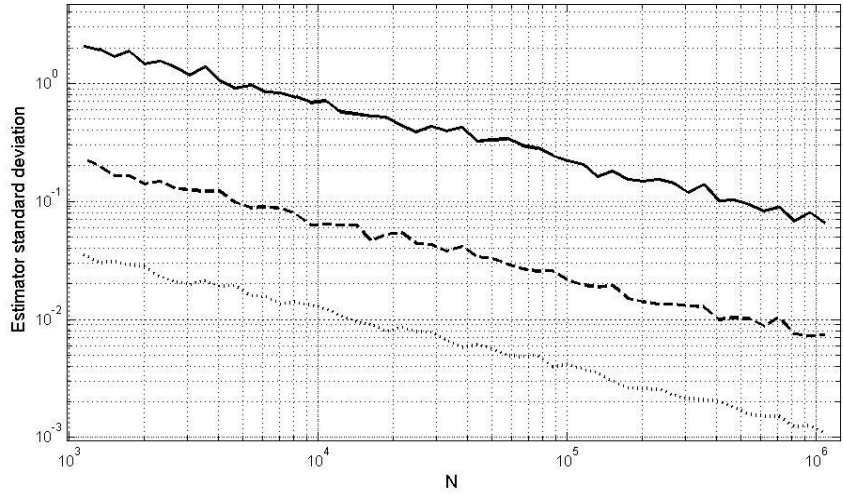

**Figure 6**

We can see that also in this extreme case the matched filter estimator works fine (the other two give much poorer results).

In figures 7 and 8 there is the case of absence of disturbances:

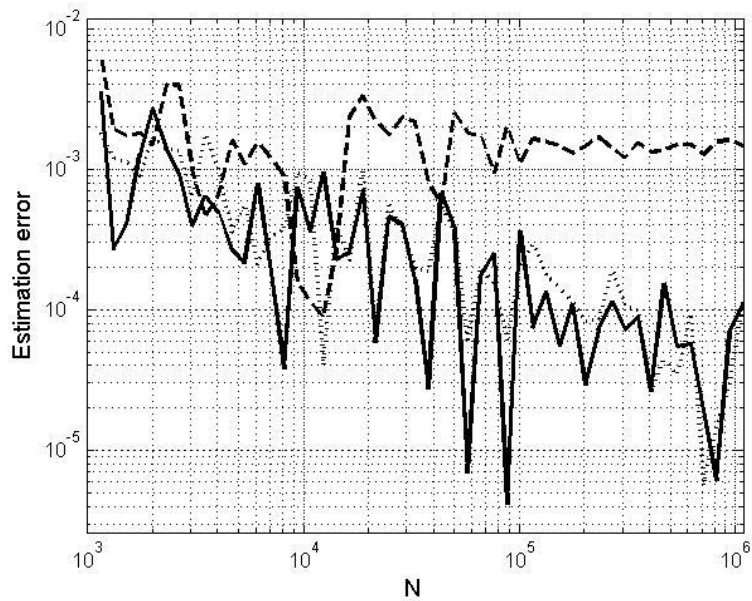

**Figure 7**

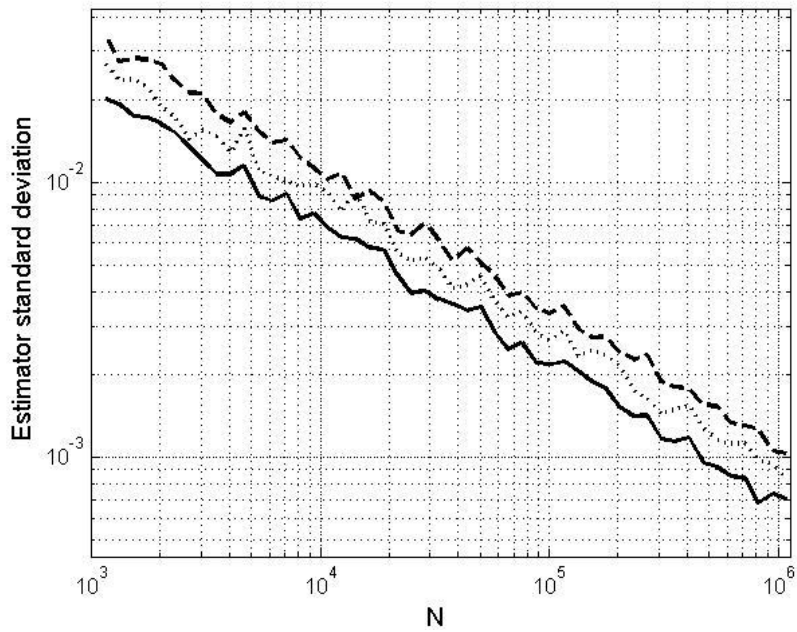

**Figure 8**

we can see that in this case, as we expected, the best results are from the classical estimator, but the results of the matched filter estimator are not much worse and are better than the log estimator.

Note that the matched filter estimator, at the same disturbance rate, gives better results for bigger disturbances. This because these give less effect in the zone of the histogram where the non-disturbed data are predominant.

**6. The non-stationary case**

If, as it is often in the real cases, the data are non-stationary, our method can be modified, constructing a "running auto-regressive" normalized histogram, with a $\tau$ chosen depending on the non-stationarity time constant, and applying our estimators on this.
To understand the method of the "running auto-regressive histogram", let us compute many subsequent normalized sub-histograms of the $z_i$ samples (on, say, 1000 samples each) and let us indicate each of them as $h_i$. Then, chosen a constant $w = e^{-\frac{1}{\tau}} < 1$, compute

(11)
$$H_i' = h_i + w \cdot H_{i-1}'$$
$$k_i = 1 + w \cdot k_{i-1}$$
$$H_i = \frac{H_i'}{k_i}$$

$H_i$ is the autoregressive histogram. The $\tau$ is the "memory" of the AR estimator.

If it is not clear what is the best $\tau$, we use banks of estimators, with different parameters (number of samples for each sub-histogram and value of $\tau$).

In figure 9 there is the results for a simulated non-stationary process, disturbed by Poissonian big pulses:

Robust estimation for Gaussian process parameters

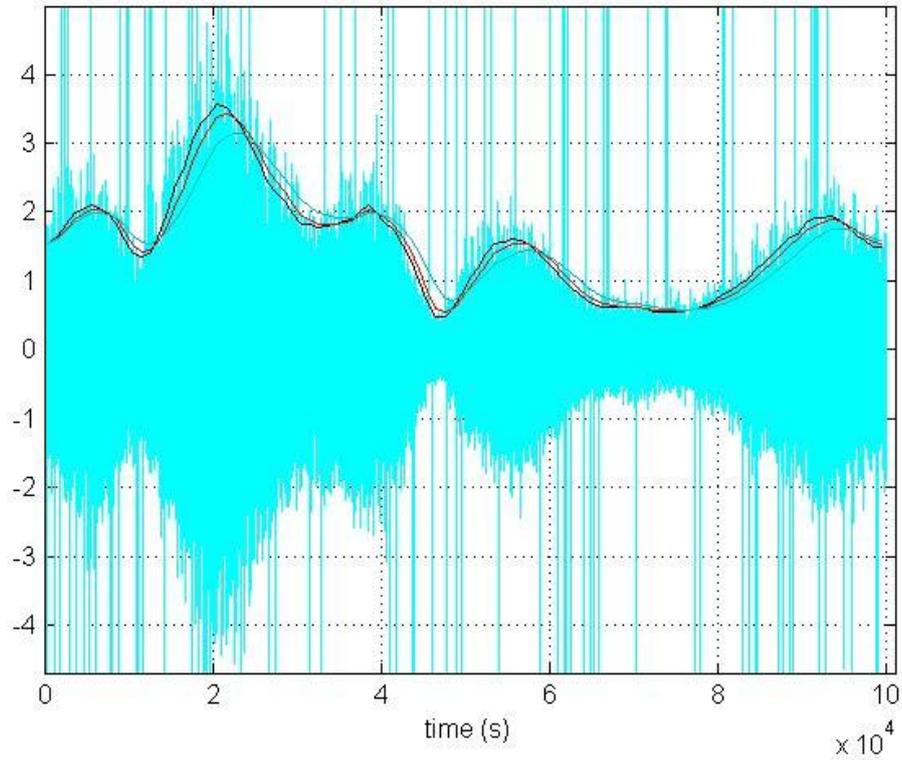

**Figure 9**

In the figure 10 there is a the disturbed process (the non-stationarity time scale is about 10000 s and the sampling time is 1 s); there are three estimations of the varying standard deviation, multiplied by 2.5, in order to follow the visible crests. In figure 10 there are the three estimates of the standard deviation (the sub-histograms are made on 1000 samples and the estimation taus are 1000, 2000 and 4000 seconds):

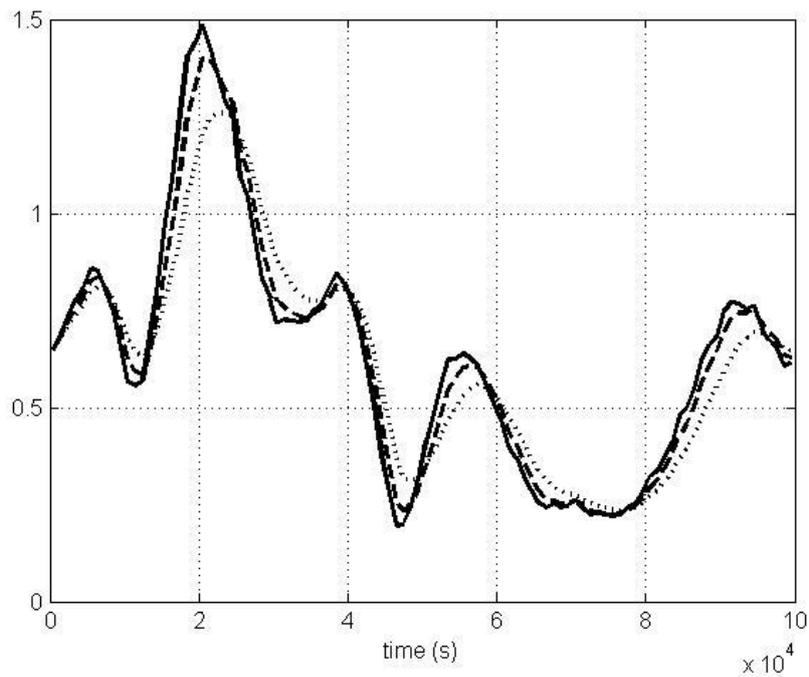

**Figure 10**

## 7 Application to the spectra

A similar idea can be applied to estimate the mean value of an exponential distributed process with big and short disturbances added, as is for the single periodogram power spectrum estimations.
In this case we can take simply z=log(x) and estimate the mean value $\mu$ of the x process.
The distribution of z (that is the shape of the matched filter) is in this case

(12) $$f(z) = \exp(z - \log(\mu)) \cdot \exp\left(-e^{z - \log(\mu)}\right)$$

So we can apply the method of the matched filter with this template.

Alternatively (but less robust of the matched filter), because the expected value of (12) is

(13) $$E[z] = \log(\mu) - \frac{1}{\sqrt{3}}$$

we can use the following estimator of the mean value

(14) $$\hat{\mu} = \exp\left(\bar{z} + \frac{1}{\sqrt{3}}\right)$$

where $\bar{z}$, as in the Gaussian case, is the average value of the $z_i$ samples. In figure 11 is the graph of the (12) function together with the (3) (dashed):

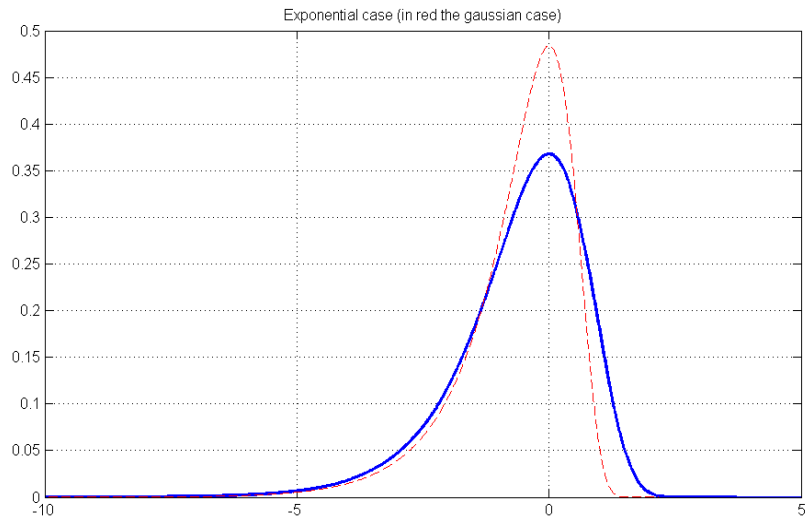

Figure 11

## 8  Conclusions

We use this adaptive matched filter method for the determination of the thresholds used to detect (and eliminate) the big disturbance event in the gravitational wave data (Astone, 2005). It is also used to defining the varying sensitivity in order to compute the Wiener filter used for the detection of the periodic sources.